\documentclass[12pt]{iopart}

\usepackage{iopams}  
\usepackage{graphicx}

\begin{document}
\title{A weak-value model for virtual particles supplying the electric current in graphene: the minimal conductivity and the Schwinger mechanism}

\author{Kazuhiro Yokota and Nobuyuki Imoto}
\address{Department of Materials Engineering Science,
Graduate School of Engineering Science, Osaka University,
Toyonaka, Osaka 560-8531, Japan}
\ead{yokota@qi.mp.es.osaka-u.ac.jp}
\date{\today}
\begin{abstract}
We propose a model for the electric current in graphene in which electric carriers are supplied by virtual particles allowed by the uncertainty relations.
The process to make a virtual particle real is described by a weak value of a group velocity: the velocity is requisite for the electric field to give the virtual particle the appropriate changes of both energy and momentum.
With the weak value, we approximately estimate the electric current, considering the ballistic transport of the electric carriers.
The current shows the quasi-Ohmic with the minimal conductivity of the order of $e^2/h$ per channel.
Crossing a certain ballistic time scale, it is brought to obey the Schwinger mechanism.
\end{abstract}

\maketitle

\section{Introduction}
\label{sec:introduction}
Graphene is fascinating material due to its applicability for electronic devices and its physical properties are also attractive in fundamental physics\cite{gra1}.
In a single layer graphene, the low energy excitation of a quasi particle can be well described by the 2+1 dimensional massless Dirac equation.
With Pauli matrices {$\hat{\sigma}_i$}, the Hamiltonian can be represented by
\begin{eqnarray}
\hat{H}=v_f(\hat{\sigma}_x \hat{p}_x +\hat{\sigma}_y \hat{p}_y), \label{eq:hamiltonian}
\end{eqnarray}
where $v_f$ is the Fermi velocity, which corresponds to the velocity of light $c$.
The absolute velocity of a particle always takes $v_f$ like a photon.
Consequently, graphene can be a tool for demonstrating relativistic phenomena like Klein's paradox\cite{gra_Klein} and Schwinger mechanism\cite{Schwinger_gra1, Schwinger_gra2}, which must be confirmed in the electrodynamics.

On the electric current in graphene, when the chemical potential and the temperature were zero, the minimal conductivity was experimentally found, of which the order was $e^2/h$ per channel (per valley and per spin)\cite{gra_cond}: the electric current $j$ shows the linear response on the electric field $\varepsilon$ as $j\sim(e^2/h)\varepsilon$, which is called the quasi-Ohmic.
Theoretical works have succeeded in obtaining the minimal conductivity, using the linear response theory\cite{gra_o1, gra_o2, gra_o3, gra_o4, gra_o5, gra_o6}, Landauer formula\cite{gra_L} and the dynamical approach\cite{gra_bal2, gra_bal3}.
Although their results show subtle different values like $e^2/(\pi h)$, there is a consensus on the minimal conductivity of the order of $e^2/h$ per channel.
Furthermore, it was also predicted that, as the electric field is stronger, so the electric current is beyond the linear response to the electric field as $j\propto\varepsilon^{3/2}$, which is owing to the Schwinger mechanism\cite{Schwinger_gra1, Schwinger_gra2, gra_bal2, gra_bal3}.
Schwinger mechanism originally represents a particle-antiparticle creation from a vacuum in a uniform electric field\cite{Schwinger1}, while a hole substitutes for an antiparticle in graphene.
The electric current can be considered as the ballistic transport of charges, since the ballistic time is long in graphene: the physical behavior can be assigned by the ballistic time.
In fact, with the ballistic time $t_{bal}$, the electric current by the Schwinger mechanism is approximately given by $j\sim en(t_{bal})v_f$, where $n(t_{bal})$ represents the density of the electric carriers (charges).
$n(t_{bal})$ can be derived from the 2+1 dimensional massless ($m=0$) pair creation rate of the Schwinger mechanism\cite{Schwinger_gra1, Schwinger2},
\begin{eqnarray} 
\frac{dn}{dt} &=& \frac{e^{3/2}\varepsilon^{3/2}}{4\pi^2\hbar^{3/2}c^{1/2}}{\rm exp}\left(-\frac{\pi m^2c^3 }{e\varepsilon\hbar}\right) \\ 
&=& \frac{e^{3/2}\varepsilon^{3/2}}{4\pi^2\hbar^{3/2}v_f^{1/2}} \ \ \ (m=0, c=v_f). \label{eq:Sch_rate}
\end{eqnarray}

Whether the electric current shows the quasi-Ohmic or the Schwinger mechanism, on first glance, it is surprising that graphene is capable of leading a current.
There is no electric carrier when the chemical potential and the temperature are zero: the density of states is proportional to the absolute value of the energy, $|E|$ \cite{gra1}.
Consequently, there must be two processes for the electric current: creation and acceleration (or reorientation
\footnote{
Note that the absolute velocity of a particle must be $v_f$.
Then, `reorientation' will be more precise.
}
).
If the ballistic time $t_{bal}$ is long enough, an electric carrier can be accelerated to the direction of the electric field after the creation.
When the electric current is mostly composed of the creation processes, it behaves as the quasi-Ohmic.
On the other hand, as the contribution of the acceleration processes surpasses the previous one, it shows the Schwinger mechanism, in which all the electric carriers are effectively in the direction of the electric field with the velocity of $v_f$, i.e. $j\sim en(t_{bal})v_f$.
The time scale of the ballistic time for their crossover is given by
\begin{eqnarray}
t_{c}=\sqrt{\frac{\hbar}{e\varepsilon v_f}}. \label{eq:t_bal}
\end{eqnarray}
As the electric field is stronger, this crossover time becomes smaller and the Schwinger mechanism will appear.
In earlier studies\cite{Schwinger_gra2, gra_bal2, gra_bal3}, it was found that, while the quasi-Ohmic current could be obtained with the perturbation on the electric field, the electric current showed the Schwinger mechanism at last in which the perturbative treatment failed beyond the time scale (\ref{eq:t_bal}).

In \cite{step}, we showed the case when a group velocity of a quantum particle was given by a weak value in the 1+1 dimensional Dirac equation, 
which was applied to a transmission through a supercritical step potential.
In this paper, we show that this weak-value formalism is also valid for describing the creation process of an electric carrier in graphene.
In fact, it has been pointed out that a weak value is useful for a description of a localized event like a pair creation\cite{W_D}.

To begin with, a weak value is introduced as a result of weak measurements: using weak measurements, we can extract a physical value on an observable without disturbing a quantum system to be measured\cite{W1, W2}.
Actually, direct observations of quantum systems have been performed\cite{W_ob1, W_ob2, W_ob3}.
In optical physics, the signal amplification effect of weak measurements has also been studied for high sensitive measurements like observations of the Hall effect\cite{W_s1}, a beam deflection\cite{W_s2}, a phase shift\cite{W_s3, W_s4}, and the Kerr nonlinearities\cite{W_s5}, including the theoretical researches\cite{W_s6, W_s7, W_s8, W_s9}.
In solid systems, such effect has been used for a charge sensing\cite{W_so1} and an atomic spontaneous emission\cite{W_so2}.
In addition to the applications as mentioned above, weak measurements have offered interesting approaches for the foundation of quantum mechanics like quantum paradoxes\cite{W_f1, W_f2, W_f3, W_f4, W_f5} and the violation of the Leggett-Garg inequality\cite{W_Leggett1, W_Leggett2, W_Leggett3, W_Leggett4}.
Apart from weak measurements, a weak value has been useful for explaining quantum phenomena\cite{step, W_D, W_en, W_tun, W_tun2, W_sup1}.
Then, the significance of a weak value itself has also been discussed in the context of a measurement\cite{W_sig1, W_sig2, W_sig3, W_sig4} and the validity for a description of quantum mechanics\cite{W_Leggett4, W_sig5, W_sig6, W_sig7, W_sig8, W_sig9}.

Our result shows one of the interesting cases in which a weak value emerges as a real value of a physical quantity like \cite{step}.
In addition, aside from an issue of a weak value, it also gives a new insight into graphene in the sense that the creation process is related to virtual particles allowed by the uncertainty relations.
We focus on just the creation process and do not care the acceleration one.
Nevertheless, it is enough for approximating the electric current for each mechanism, the quasi-Ohmic and the Schwinger mechanism.

We show that a weak value may also appear as a group velocity even in the 2+1 dimensional massless Dirac equation here\cite{step}.
According to equation (\ref{eq:hamiltonian}), a plane wave with a (positive or negative) energy $\pm E$ and a momentum $p=(p_x,p_y)$ can be described as follows,
\begin{eqnarray}
\frac{1}{\sqrt 2}
\left[ 
\begin{array}{c}
e^{-i\theta/2} \\
\pm e^{i\theta/2}\\
\end{array}
\right]e^{\frac{i}{\hbar}(p_x x+p_y y)} = |\pm E\rangle \psi_{p_x,p_y}(x,y), \label{eq:eigenstate}
\end{eqnarray}
where $\theta={\rm Arctan}(p_y/p_x)$ shows the direction of the momentum.
They satisfy the energy-momentum relation, $E^2=v_f^2p^2=v_f^2p_x^2+v_f^2p_y^2$.
$|\pm E\rangle$ is independent of $x$ and $y$, which is called the chirality.
The dependent part $\psi_{p_x,p_y}(x,y)$ is called the space part.
Consider the case that a chirality prepared in the initial state $|E\rangle$ is finally found in $|E'\rangle$, which is referred to as the preselection in $|E\rangle$ and the postselection in $|E'\rangle$.
When $t$ is small enough, the time evolution of the space part is given as follows,
\begin{eqnarray}
& &\langle E'| e^{-\frac{i}{\hbar}v_f(\hat{\sigma}_x\hat{p}_x+\hat{\sigma}_y\hat{p}_y)t}|E\rangle \psi_{p_x,p_y}(x,y)  \label{eq:t_ev1} \\
&\sim& \langle E'|E\rangle e^{-\frac{i}{\hbar}v_f\langle\hat{\sigma}_x\rangle _{\bf w}\hat{p}_xt}e^{-\frac{i}{\hbar}v_f\langle\hat{\sigma}_y\rangle_{\bf w}\hat{p}_yt}
\psi_{p_x,p_y}(x,y) \ \ \ (t \ \sim \ 0) \label{eq:t_ev_app1} \\
&\sim& \langle E'|E\rangle \psi_{p_x,p_y}(x-v_f\langle\hat{\sigma}_x\rangle_{\bf w}t,y-v_f\langle\hat{\sigma}_y\rangle_{\bf w} t)\ \ \ (t \ \sim \ 0), \label{eq:WVshift1}
\end{eqnarray}
where $\langle\hat{\sigma}_x\rangle_{\bf w}$ is a weak value,
\begin{eqnarray}
\langle\hat{\sigma}_x\rangle_{\bf w}=\frac{\langle E'|\hat{\sigma}_x|E\rangle}{\langle E'|E\rangle}, \label{eq:weakvalue}
\end{eqnarray}
and the weak value of $\hat{\sigma}_y$ is given in a similar way.
$v_f\langle \hat{\sigma}_x\rangle_{\bf w}$ and $v_f\langle \hat{\sigma}_y\rangle_{\bf w}$ correspond to the (group) velocities in $x$ and $y$ directions respectively.
In fact, without the postselection, they give $v_f\langle \hat{\sigma}_x\rangle=p_xv_f^2/E=v_f{\rm cos}\theta$ and $v_f\langle \hat{\sigma}_y\rangle=p_yv_f^2/E=v_f{\rm sin}\theta$.
If $|E'\rangle$ represents the chirality of the eigenstate of the energy $E'$ and the momentum $p'=(p'_x,p'_y)$, the weak values are given as follows,
\begin{eqnarray}
\langle\hat{\sigma}_x\rangle_{\bf w} &=& \frac{{\rm sin}[(\theta+\theta ')/2]}{{\rm sin}[(\theta-\theta ')/2]}  \label{eq:wv_x_gen} \\
\langle\hat{\sigma}_y\rangle_{\bf w} &=& -\frac{{\rm cos}[(\theta+\theta ')/2]}{{\rm sin}[(\theta-\theta ')/2]} , \label{eq:wv_y_gen}
\end{eqnarray}
where $\theta '={\rm Arctan}(p'_y/p'_x)$.
Although a weak value is generally a complex number as shown in equation (\ref{eq:weakvalue}), it is always real number as far as considering energy eigenstates in our case. That is why we treat a weak value as a real number hereafter.

In the next section, considering a transition between energy eigenstates, we try to describe a creation process in graphene by a pre-postselection of a chirality.
We show that the weal value of a group velocity (\ref{eq:weakvalue}) is requisite for the electric field to yield the changes of both the energy and the momentum appropriately for such transition.
In section \ref{sec:virtual}, we assume that the creation process for an electric carrier is triggered off by a virtual particle, which is allowed by the uncertainty relations. 
The weak-value formalism for describing a time evolution like (\ref{eq:WVshift1}) is justified for such virtual particles, although $t$ is not always $\sim 0$.
In section \ref{sec:transport}, we approximately estimate the electric current in graphene, using a weak value of a group velocity.
It is shown that, crossing the time scale (\ref{eq:t_bal}), the current flows in the different manners, namely, the quasi-Ohmic and the Schwinger mechanism.
Section \ref{sec:conclusion} is devoted to our conclusion.

\section{A transition for creating an electric carrier in graphene}
\label{sec:trans}

\begin{figure}
  \begin{center}
	 \includegraphics[scale=0.7]{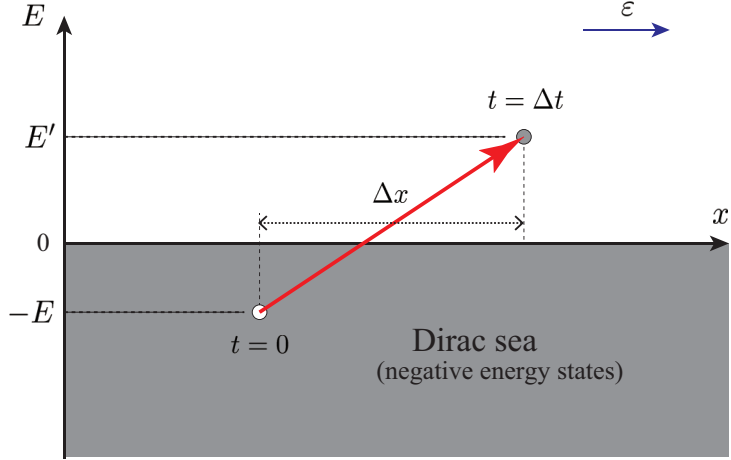}
  \end{center}
  \caption{A transition from a negative energy to a positive one.
During the transition, a particle moves into $\Delta x$, which takes the time $\Delta t$.
The vacancy of a particle in the Dirac sea corresponds to a hole, that is, a transition represents a creation of a particle-hole pair.
}
  \label{fig:transition}
\end{figure}
First of all, there must be a creation process of an electric carrier in graphene so as to be capable of leading an electric current.
An electric carrier will be supplied by creating a particle-hole pair, which is represented by a transition of a particle in the Dirac sea (valence band) to the vacuum (conduction band).
For this purpose, we consider a transition as shown in figure \ref{fig:transition}, supposing that the electric field $\varepsilon$ is in $x$ direction.
Initially, a particle in the Dirac sea has a negative (kinetic) energy $-E$ and a momentum $(-p_x,p_y)$, where $E, p_x\ge 0$.
By the electric field, the particle might change to the one with a positive (kinetic) energy $E'$ and a momentum $(p'_x,p_y)$, where $E', p'_x\ge 0$.
The momentum in $y$ direction does not change, because the electric field is zero in this direction.
The signs of $p_x$ and $p'_x$ provide $+x$ velocities, which is in the direction of the electric field, because they give the group velocities $(-p_xv_f^2)/(-E)\ge 0$ and $p'_xv_f^2/E'\ge 0$ respectively.
As an energy eigenstate can be specified by a chirality as shown in equation (\ref{eq:eigenstate}), this transition process can be described by the pre-postselection on the chirality, $|E\rangle$ and $|E'\rangle$.
Like equation (\ref{eq:WVshift1}), when the time $t$ is very small, the time evolution of the space part can be approximately expressed as follows,
\begin{eqnarray}
& &\langle E'| e^{-\frac{i}{\hbar}(v_f(\hat{\sigma}_x\hat{p}_x+\hat{\sigma}_y\hat{p}_y)-e\varepsilon x)t}|E\rangle \psi_{p_x,p_y}(x,y)  \label{eq:t_ev2} \\
&\sim& \langle E'|E\rangle e^{-\frac{i}{\hbar}v_f\langle\hat{\sigma}_x\rangle _{\bf w}\hat{p}_xt}e^{-\frac{i}{\hbar}v_f\langle\hat{\sigma}_y\rangle_{\bf w}\hat{p}_yt}e^{\frac{i}{\hbar}e\varepsilon x t}\psi_{p_x,p_y}(x,y)  \ \ \ (t \ \sim \ 0) \label{eq:t_ev_app2} \\
&\sim& \langle E'|E\rangle \psi_{p_x+e\varepsilon t,p_y}(x-v_f\langle\hat{\sigma}_x\rangle_{\bf w}t,y-v_f\langle\hat{\sigma}_y\rangle_{\bf w} t) \ \ \ (t \sim \ 0).  \label{eq:WVshift2}
\end{eqnarray}
This is different from equation (\ref{eq:WVshift1}) in the respect that there is a momentum shift for $p_x$ due to the electric field \cite{step}.
At this stage, however, it is not clear whether this weak-value formalism is valid, as we have not yet verified that the time is small enough for this approximation.
The validity of the approximation will be discussed in the next section.
At any rate, the weak values of the group velocities in $x$ and $y$ directions can be respectively defined by equation (\ref{eq:wv_x_gen}) and equation (\ref{eq:wv_y_gen}), with $\theta={\rm Arctan}(p_y/(-p_x))$ and $\theta '={\rm Arctan}(p_y/p'_x)$.
These group velocities represent the velocities driven by the transitions.
Because the electric field is zero in $y$ direction, the group velocity in this direction should be zero, namely,
\begin{eqnarray}
\langle \hat{\sigma}_y\rangle_{\bf w}=0,
\end{eqnarray}
equivalently,
\begin{eqnarray}
\theta'+\theta=\pm\pi \ \ {\rm i.e.}, \ \ p_x'=p_x \ {\rm and} \ E'=E. \label{eq:select_trans}
\end{eqnarray}
If not, particles seem to accomplish transitions with zero electric field and the velocity, which causes the current, emerges in $y$ direction.
Equation (\ref{eq:select_trans}) shows that a transition is selective\cite{gra_np}: a particle with a negative energy $-E$ and a momentum $(-p_x, p_y)$ may turn out one with $E$ and $(p_x, p_y)$.
In this case, $\langle \hat{\sigma}_x\rangle_{\bf w}$ is given by
\begin{eqnarray}
\langle \hat{\sigma}_x\rangle_{\bf w} = \frac{\sqrt{p_x^2+p_y^2}}{p_x}=\frac{1}{{\rm cos}\theta}  >1, \label{eq:WV_x}
\end{eqnarray}
which is a strange value, that is, the corresponding group velocity $v_f\langle\hat{\sigma}_x\rangle_{\bf w}$ is more than $v_f$\cite{W_sup1, W_sup2, W_sup3}.
The appearance of such strange weak value agrees with the result of \cite{step} due to a transition from a negative energy state to a positive one.

To clarify the meaning of this strange velocity, we consider the inside details of the transition process.
During the transition, the changes of the energy and the momentum are $2E(=E-(-E))\equiv\Delta E$ and  $2p_x(=p_x-(-p_x))\equiv\Delta p_x$ respectively.
The force $e\varepsilon$ by the electric field acts on a particle in $x$ direction.
Define $\Delta x$ as the moving distance for the duration of the transition.
As the energy change is equivalent to the work done by the electric field, it satisfies,
\begin{eqnarray}
\Delta E=e\varepsilon\Delta x. \label{eq:e_change}
\end{eqnarray}
With the time needed for the transition $\Delta t$, we also obtain,
\begin{eqnarray}
\Delta p_x = e\varepsilon\Delta t, \label{eq:p_change}
\end{eqnarray}
because of the equivalence between the momentum change and the impulse.
Then, we can define the average (group) velocity $v_g$ during the transition process as follows,
\begin{eqnarray}
v_g &\equiv& \frac{\Delta x}{\Delta t}  \\
&=&\frac{\Delta E}{\Delta p_x}=\frac{2E}{2p_x}=\frac{v_f\sqrt{p_x^2+p_y^2}}{p_x}.  \label{eq:ave_v}
\end{eqnarray}
From equations (\ref{eq:WV_x}) and (\ref{eq:ave_v}), we can find,
\begin{eqnarray}
v_g=v_f\langle \hat{\sigma}_x\rangle_{\bf w},
\end{eqnarray}
which shows that, in fact, the weak value of the group velocity (\ref{eq:WV_x}) is requisite to satisfy the energy change (\ref{eq:e_change}) and the momentum change (\ref{eq:p_change}) simultaneously.

\begin{figure}
  \begin{center}
	 \includegraphics[scale=0.7]{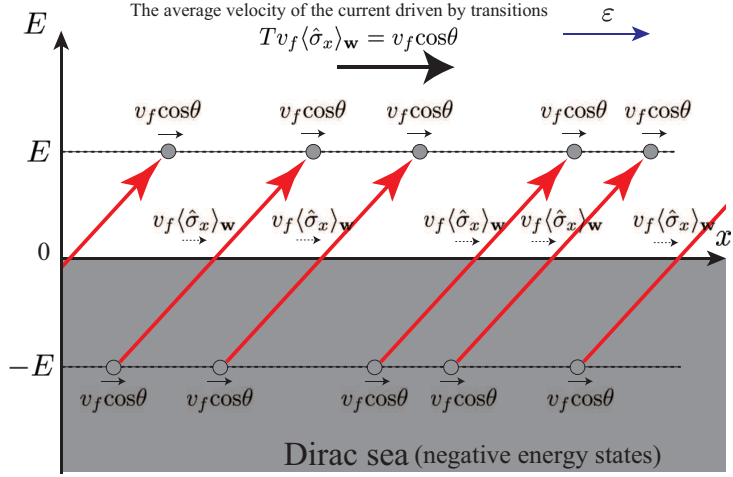}
  \end{center}
  \caption{The current due to transitions from a negative energy level $-E$ to the positive one $E$, except for the accelerations after the transitions.
A particle in $-E$ ($E$) has the group velocity $v_f{\rm cos}\theta$ just before (after) a transition.
The current driven by the creations is composed of such particles with a homogeneous density. 
Note that the current in the negative energy states corresponds to the flux of holes in the opposite direction.
The entire flux shows the current of the charges in $+x$ direction.
As a particle has the group velocity $v_f\langle\hat{\sigma}_x\rangle_{\bf w}$ during a transition,
a transition should occur with a probability $T$ to agree with the average velocity of the flux as shown in equation (\ref{eq:trans_ave}).
}
  \label{fig:trans_prob}
\end{figure}

Using a weak value, we can also estimate a probability of occurring a transition.
In a specific postselection, a weak value may take a strange value lying outside of the spectra of eigenvalues.
However, the average value should be within the conventional range of value in considering all the possible postselection.
In our case, when it succeeds in postselecting a chirality by a positive energy eigenstate $|E\rangle$, $v_f\langle\hat{\sigma}_x\rangle_{\bf w}$ is more than $v_f$ as shown in equation (\ref{eq:WV_x}).
Such weak value yields a group velocity of a current due to the transition process, because a transition corresponds to a creation of a particle-hole pair, namely, a carrier.
Note that the current does not contain the effect of the process after the creation, i.e. the acceleration.
Without a transition, it brings about zero group velocity and does not contribute to generating a current, because a particle keeps in the Dirac sea as before.
Including such non-transition particles, the average velocity should be conventionally less than $v_f$, by which the net velocity of the current is given.
After all, the current does not flow beyond $v_f$ like superluminal velocity: it never occurs as a strange physical phenomenon with a strange value of a physical quantity as the whole.
For this reason, all the particle do not transmit to a positive energy, and such transition happens with some probability $T$.
In \cite{step}, we could actually estimate the transmission probability for a step potential by making a weal value of a group velocity at the step consistent with an average velocity of the flux outside the step.
In a similar way, we can obtain a transition probability $T$ as shown in figure \ref{fig:trans_prob}.
The number of transition particles in a positive energy eigenstate $|E\rangle$ is equivalent to the one in a negative energy eigenstate $|-E\rangle$, which corresponds to the holes.
In addition, the group velocities of particles just before and just after a transition, namely, the group velocities in $|E\rangle$ and $|-E\rangle$ are the same as $v_f{\rm cos}\theta(=p_xv_f^2/E=(-p_xv_f^2)/(-E))$.
Consequently, the average velocity of the current driven by the transition should be also $v_f{\rm cos}\theta$.
On the other hand, as we have mentioned, the transition itself generates the group velocity $v_f\langle\hat{\sigma}_x\rangle_{\bf w}$ more than $v_f{\rm cos}\theta$.
If the transition probability is given by $T$, the average velocity $Tv_f\langle\hat{\sigma}_x\rangle_{\bf w}$ has to satisfy
\begin{eqnarray}
Tv_f\langle\hat{\sigma}_x\rangle_{\bf w}=v_f{\rm cos}\theta. \label{eq:trans_ave}
\end{eqnarray}
From this equation, we can find the transition probability as follows,
\begin{eqnarray}
T(p_x,p_y)={\rm cos}^2\theta=\frac{p_x^2}{p_x^2+p_y^2}. \label{eq:trans_prob}
\end{eqnarray}
This transition probability is the same as the transmission probability for the step potential shown in \cite{gra_np}, in which n-p junction in graphene is treated.
This agreement is plausible, as we have derived the `transition' probability in the same manner of the `transmission' probability for the step potential like \cite{step}.
They have a common point that the electric field brings about the process (transition or transmission), which generates and determines the velocity of the current.

We have discussed a transition between energy eigenstates by an electric field.
A creation of an electric carrier in graphene should be described by this picture.
In the next section, we consider how much energy states can participate in such transitions for creating electric carriers.

\section{An electric carrier and the uncertainty relations}
\label{sec:virtual}
We assume that a transition for creating an electric carrier is triggered off by fluctuations allowed by the uncertainty relations.
According to the uncertainty relation between energy and time, 
\begin{eqnarray}
\delta E \delta t \sim \hbar, \label{eq:unET}
\end{eqnarray}
an energy fluctuation $\delta E$ can occur during a time $\delta t$, which means that a virtual particle with the energy $\delta E$ can exist during the lifetime $\delta t$.
In a similar way, a virtual particle with a momentum $\delta p_x$ can be considered within a space $\delta x$ in $x$ direction, and they satisfy the uncertainty relation,
\begin{eqnarray}
\delta x\delta p_x \sim \hbar. \label{eq:unXP}
\end{eqnarray}
In our case, the virtual particle can correspond to a virtual transition like figure \ref{fig:transition}, and has the energy $2E\equiv\delta E$ and the momentum $2p_x\equiv\delta p_x$.
Note that we are concerned about the case in which the momentum change in $y$ direction is zero due to zero electric field and do not have to care the fluctuation of the momentum in this direction.
Such virtual transition with the energy $\delta E$ and the momentum $\delta p_x$, however, is not always consistent with a real particle, because it does not always satisfy the appropriate energy-momentum relation, $\delta E^2=v_f^2(\delta p_x^2+p_y^2)$: 
a virtual particle satisfy the uncertainty relations (\ref{eq:unET}) and (\ref{eq:unXP}) independently.
In fact, to make a virtual particle contribute to the electric current as a real one would, the electric field must satisfy (\ref{eq:e_change}) and (\ref{eq:p_change}) simultaneously, which is, as we mentioned before, accomplished by the weak value of the group velocity (\ref{eq:WV_x}).
In other words, a virtual particle gives the electric field a chance to do the work and the impulse, by which we mean the electric current is able to pass in graphene. 
Within the lifetime $\delta t=\hbar/2E=\hbar/(2v_f\sqrt{p_x^2+p_y^2})$, the electric field must achieve the impulse $2p_x=\Delta p_x$, which takes the time $\Delta t$ (see
 equation (\ref{eq:p_change})), namely,
 \begin{eqnarray}
 \Delta t \le \delta t,  \label{eq:t_t}
 \end{eqnarray}
from which we can obtain
\begin{eqnarray}
p_x^2(p_x^2+p_y^2)\le\frac{e^2\varepsilon^2\hbar^2}{16v_f^2}. \label{eq:virtual}
\end{eqnarray}
(\ref{eq:virtual}) assigns the energy states which may contribute to electric carriers via virtual particles.
The same result can be derived from the relation between the work and the space instead of the impulse and the time: 
according to equation (\ref{eq:e_change}), the work $2E=\Delta E$ needs the space $\Delta x$, which should be smaller than the fluctuation $\delta x=\hbar/2p_x$ as follows,
\begin{eqnarray}
\Delta x \le \delta x.  \label{eq:x_x}
\end{eqnarray}
Satisfying (\ref{eq:e_change}) and (\ref{eq:p_change}) simultaneously, the weak value makes the uncertainty relations (\ref{eq:unET}) and (\ref{eq:unXP}) equivalent in the sense that it selects a real particle from virtual particles in the independent uncertainty relations (\ref{eq:unET}) and (\ref{eq:unXP}).
As a result, it is plausible that (\ref{eq:t_t}) and (\ref{eq:x_x}) derive the same result (\ref{eq:virtual}), because the weak value satisfies the appropriate changes of both the energy and the momentum.

So far, we have proceeded to a discussion as if the approximation of the weak-value formalism, equation (\ref{eq:WVshift2}), is valid and a velocity of a particle during a transition is given by a weak value.
As follows, we verify that this approximation is adequate as far as the above-mentioned transition stemming from a virtual particle.
Expanding on $t$, we can describe equation (\ref{eq:t_ev2}) as follows,
\begin{eqnarray}
& &\langle E'| e^{-\frac{i}{\hbar}(v_f(\hat{\sigma}_x\hat{p}_x+\hat{\sigma}_y\hat{p}_y)-e\varepsilon x)t}|E\rangle \psi_{p_x,p_y}(x,y) \nonumber \\
&=& \langle E'|E\rangle\sum_{k=0}^{\infty}\frac{1}{k!}\Bigl(-\frac{i}{\hbar}t\Bigr)^k\frac{\langle E'|(v_f(\hat{\sigma}_x\hat{p}_x+\hat{\sigma}_y\hat{p}_y)-e\varepsilon x)^k|E\rangle}{\langle E'|E\rangle} \psi_{p_x,p_y}(x,y)
\end{eqnarray}
This equation coincides with equation (\ref{eq:t_ev_app2}) by the first order of $t$, which is given by
\begin{eqnarray}
& & \langle E'|E\rangle e^{-\frac{i}{\hbar}v_f\langle\hat{\sigma}_x\rangle _{\bf w}\hat{p}_xt}e^{-\frac{i}{\hbar}v_f\langle\hat{\sigma}_y\rangle_{\bf w}\hat{p}_yt}e^{\frac{i}{\hbar}e\varepsilon x t}\psi_{p_x,p_y}(x,y) \nonumber \\
&=& \langle E'|E\rangle\sum_{k=0}^{\infty}\frac{1}{k!}\Bigl(-\frac{i}{\hbar}v_f\langle\hat{\sigma}_x\rangle_{\bf w}\hat{p}_xt\Bigr)^k \nonumber \\
& & \ \ \ \ \ \ \sum_{k'=0}^{\infty}\frac{1}{k'!}\Bigl(-\frac{i}{\hbar}v_f\langle\hat{\sigma}_y\rangle_{\bf w}\hat{p}_yt\Bigr)^{k'}
\sum_{k''=0}^{\infty}\frac{1}{k''!}\Bigl(\frac{i}{\hbar}e\varepsilon xt\Bigr)^{k''}\psi_{p_x,p_y}(x,y).     \label{eq:t_ev_app2_exp}
\end{eqnarray} 
Consequently, equation (\ref{eq:t_ev2}) can be approximated to equation (\ref{eq:t_ev_app2}), when the terms of $O(t^k)$ $(k\geq 2)$ can be neglected, which should be properly satisfied in $t\sim 0$.
If it satisfies not $t\sim 0$ but that the higher terms of $O(t^k)$ $(k\geq 2)$ are smaller than the first one, however, this approximation will also stand for rough estimations.
In our case, we can obtain all what we need to valuate the higher terms: 
$\langle\hat{\sigma}_x\rangle_{\bf w} = O(\Delta E/\Delta p_x)/v_f$,
$\langle\hat{\sigma}_y\rangle_{\bf w} = 0$,
$\langle\hat{\sigma}_z\rangle_{\bf w} = O(\langle\hat{\sigma}_x\rangle_{\bf w})$,
$e\varepsilon = O(\Delta p_x/\Delta t)$,
$2p_x=O(\Delta p_x)$, and
$t=O(\Delta t)$.
For example, one of the second terms in equation (\ref{eq:t_ev_app2_exp}) can be estimated as follows,
\begin{eqnarray}
& & \frac{1}{2!}\left(\frac{i}{\hbar}\right)^2v_f^2\langle\hat{\sigma}_x\rangle_{\bf w}^2\hat{p}_x^2t^2\psi_{p_x,p_y}(x,y) \nonumber \\
&=&  \frac{1}{2!}\left(\frac{i}{\hbar}\right)^2O\left(\frac{\Delta E^2}{\Delta p_x^2}\right)O(\Delta p_x^2) O(\Delta t^2)\psi_{p_x,p_y}(x,y)  \nonumber \\
&=&   \frac{1}{2!}\left(\frac{i}{\hbar}\right)^2O(\Delta E^2\Delta t^2)\psi_{p_x,p_y}(x,y).
\end{eqnarray}
Because of $\Delta E\Delta t<\hbar$, this term is smaller than the first one.
In a similar way, we can verify that the other higher terms are also smaller.
As a result, it is reasonable to describe the weak-value formalism for the transition starting from a virtual particle in the uncertainty relations: 
as far as the rough estimation of the electric current, we can regard the velocity of a particle as the weak value during the transition process.

\section{The ballistic transport in graphene with a weak value}
\label{sec:transport}
In graphene, the ballistic time $t_{bal}$ is long, within which we can pay no attention to the interactions with phonons, electrons, and so on.
The effect of the disorder can also be ignored.
In the spirit of Drude model, the electric current in graphene can be explained with such ballistic transport\cite{gra_bal2, gra_bal3}: 
the ballistic time $t_{bal}$, which is mostly given by $t_{bal}=L/v_f$ with the size of the graphene sample $L$, corresponds to the mean free time.
Moving the Dirac point, the net velocity appears along the electric field and brings about the current\cite{gra1}, the behavior of which is assigned by $t_{bal}$.
In our case, $t_{bal}$ should similarly participate in the electric current.
The transition time for creating an electric carrier $\Delta t$ must be smaller than $t_{bal}$, namely, $\Delta t\le t_{bal}$,
from which we can obtain,
\begin{eqnarray}
0\le p_x\le \frac{1}{2}e\varepsilon t_{bal}.   \label{eq:bal_t}
\end{eqnarray}
The distance $\Delta x$ to achieve the transition is also smaller than $L$, i.e. $\Delta x\le L$. Then, we can find,
\begin{eqnarray}
\sqrt{p_x^2+p_y^2}\le\frac{1}{2}e\varepsilon t_{bal},   \label{eq:bal_x}
\end{eqnarray}
which includes (\ref{eq:bal_t}): 
the energy states satisfying (\ref{eq:bal_x}) can actually participate in the electric current, given $t_{bal}$.
As a result, the actual electric current should consist of the energy states in both ({\ref{eq:virtual}) and (\ref{eq:bal_x}).
Note that (\ref{eq:virtual}) represents candidates for electric carriers via virtual particles.

We have not been concerned about the amount of $t_{bal}$ itself.
However, if $t_{bal}$ is small enough, the energy fluctuation $\delta E_{bal}$ may be effective on the current where $\delta E_{bal}$ is given by the uncertainty relation of $\delta E_{bal} \ t_{bal} \sim \hbar$.
With the energy fluctuation $2E\le\delta E_{bal}$ ($E=v_f\sqrt{p_x^2+p_y^2}$), it gives
\begin{eqnarray}
\sqrt{p_x^2+p_y^2}\le \frac{\hbar}{2v_ft_{bal}},   \label{eq:bal_virtual}
\end{eqnarray}
within which the energy states are involved in the energy fluctuation $\delta E_{bal}$.
Figure \ref{fig:range} represents (\ref{eq:virtual}), (\ref{eq:bal_x}), and (\ref{eq:bal_virtual}) for the various amounts of $t_{bal}$.
Clearly, they show the different features, crossing the time scale $t_{c}$ given by (\ref{eq:t_bal}).
In $t_{bal}<t_{c}$, all states to be considered are included in the energy fluctuation by the ballistic time (\ref{eq:bal_virtual}) as denoted by O, while they are divided into two regions, namely, the inside and the outside of the fluctuation (O and S) in $t_{bal}>t_{c}$.
After all, when $t_{bal}>>t_c$, the most states are out of the energy fluctuation and belong to S.
It follows that each case of $t_{bal}$ shows a different mechanism of the electric current.

\begin{figure}
  \begin{center}
	 \includegraphics[scale=0.4]{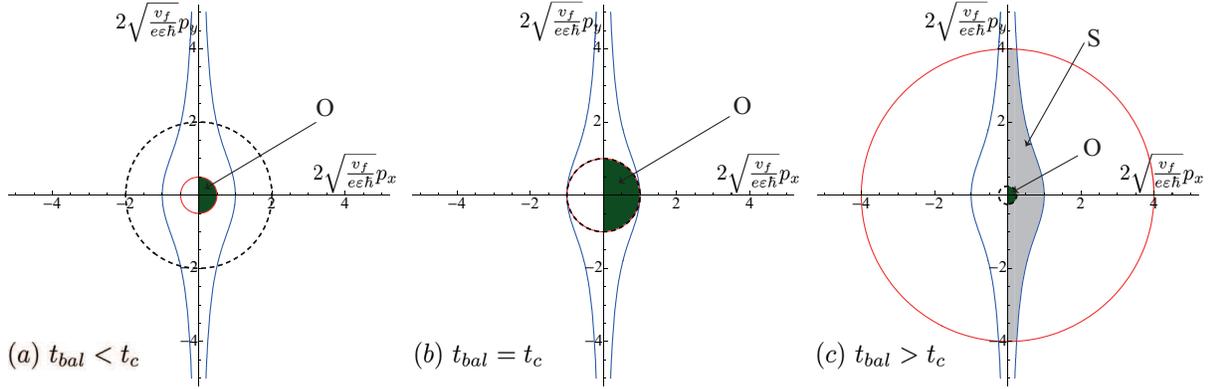}
  \end{center}
  \caption{The figures of (\ref{eq:virtual}), (\ref{eq:bal_x}), and (\ref{eq:bal_virtual}) in the various cases of $t_{bal}$: (a)$t_{bal}<t_c$, (b)$t_{bal}=t_c$, and $t_{bal}>t_c$.
Their values of $t_{bal}$ are chosen appropriately. $p_x$ and $p_y$ are also normalized suitably.
The regions surrounded by the blue curves represent (\ref{eq:virtual}), which provide candidates for electric carriers via virtual particles.
The regions assigned by the ballistic times, i.e. (\ref{eq:bal_x}) are within the red solid circles.
The energy fluctuations due to the ballistic times, which are given by (\ref{eq:bal_virtual}), are indicated by the dashed circles.
The regions satisfying both (\ref{eq:virtual}) and (\ref{eq:bal_x}), which are within both the blue curves and the red solid circles, are colored.
According as the inside or the outside of the dashed circle, they are color-coded by the dark green or the pale gray, which are referred to as O and S respectively.
Note that $p_x\ge 0$ are concerned, because the initial momentum $-p_x$ should be negative.
These colored regions provide the energy states contributing to the electric current: they assign the regions of the integrations, (\ref{eq:int_O}) and (\ref{eq:int_S}), for counting the energy states.
In (a), the colored region is utterly within the dashed circle as shown by O.
Getting the larger $t_{bal}$, the size of the red circle overtakes the dashed one in (b).
In (c), the colored regions are divided into O and S by the dashed circle, namely, the energy fluctuation by the ballistic time.
}
  \label{fig:range}
\end{figure}

The time scale of $t_{bal}<t_{c}$ corresponds to the quasi-Ohmic.
In this case, the electric current is significantly composed of the creation processes, because $t_{bal}$ is not long enough for an acceleration after a creation\cite{gra_bal2, gra_bal3}.
For this reason, although our model describes just the creation process, we can actually try to attain the result of the  quasi-Ohmic.
As shown in figure \ref{fig:range} (a), all the energy state contributing to the creation process is within the energy fluctuation by the ballistic time, (\ref{eq:bal_virtual}).
Then, for each state of an energy $E$, we can consider the number of virtual particles as $\delta E_{bal}/2E$: one state may supply more than one virtual particle.
This means that, in addition to approximating the number of contributing energy states by the uncertainty relations (\ref{eq:unET}) and (\ref{eq:unXP}), we are also trying to approximate the contribution per state, using the uncertainly relation on the ballistic time scale.
Each virtual particle provides an opportunity for the electric field to work of $2E$, which is accomplished with the time needed $\Delta t$.
Then, the work per unit time for each state is given by $2E/\Delta t$.
With the transition probability $T(p_x,p_y)$, the whole work per unit time done by the electric field can be estimated as follows,
\begin{eqnarray}
& & \frac{1}{(2\pi\hbar)^2}\int\int_{\rm O}dp_xdp_yT(p_x,p_y)\frac{2E}{\Delta t}\frac{\delta E_{bal}}{2E} \label{eq:int_O} \\
&=& \frac{1}{(2\pi\hbar)^2}\frac{e\varepsilon\hbar}{2t_{bal}}\int\int_{\rm O}dp_xdp_y\frac{p_x}{p_x^2+p_y^2}   \nonumber \\
&=& \frac{1}{(2\pi\hbar)^2}\frac{e\varepsilon\hbar}{2t_{bal}}
\int_0^{\frac{1}{2}e\varepsilon t_{bal}}dr\int_{-\pi/2}^{\pi/2}d\theta{\rm cos}\theta  \ \ \ 
(p_x=r{\rm cos}\theta, \ p_y=r{\rm sin}\theta)   \nonumber \\
&=& \frac{e^2\varepsilon^2}{4\pi h}. \label{eq:quasi_OhmicW}
\end{eqnarray}
When the electric current is proportional to the electric field as $j=\sigma\varepsilon$ with the conductivity $\sigma$, the work per unit time is given by $j\varepsilon=\sigma\varepsilon^2$.
Comparing equation (\ref{eq:quasi_OhmicW}) to $j\varepsilon$, we can find the electric current,
\begin{eqnarray}
j=\frac{e^2}{4\pi h}\varepsilon, \label{eq:quasi_OhmicJ}
\end{eqnarray}
and the conductivity,
\begin{eqnarray}
\sigma=\frac{e^2}{4\pi h}. \label{eq:quasi_OhmicS}
\end{eqnarray}
We have obtained the linearity of the electric current, namely, the quasi-Ohmic.
The estimated conductivity (\ref{eq:quasi_OhmicS}) almost accords with the minimal conductivity $\sim e^2/h$ per channel, especially the theoretical value like $e^2/(\pi h)$.
In this case, $\delta E_{bal}/2E$-fold virtual particles play roles of the electric carriers per state.
Of course, the electric field actually performs the corresponding work and impulse, which is the cause of the conductivity or the resistivity unlike Joule heat in the Ohmic.
It resembles a pi-meson taking on a nuclear force between nuclear particles, as we can approximate the mass of the pi-meson $m_\pi$ with the energy-time uncertainty relation, $\delta E_\pi\delta t_\pi=(m_\pi c^2)\delta t_\pi\sim \hbar$.
$\delta t_\pi\sim\hbar/m_\hbar c^2$ corresponds to the lifetime of the pi-meson.
Such pi-meson is effective within $c\delta t_\pi$, namely, the Compton wave length, $\hbar/m_\pi c$.
Estimating this length as the size of the atomic nuclei, $d\sim10^{-15}{\rm m}$, we can find $m_\pi c^2\sim 200{\rm MeV}$, which agrees with $m_\pi c^2\sim140{\rm MeV}$.
It is the fact that the atomic nuclei is stable due to the nuclear force with a mediation of a virtual particle of a pi-meson, which does not emerge from the nuclei.
In this sense, the virtual particle is real as far as no violation of the energy conservation, which is the same as an electric carrier for the quasi-Ohmic in graphene.

When $t_{bal}>>t_{c}$, the energy fluctuation due to the ballistic time is very small for the most states, which belong to the region S as shown in figure \ref{fig:range} (c).
For such states, the effect of the fluctuation can be neglected unlike the quasi-Ohmic case: one state supplies one particle.
While the region O provides the quasi-Ohmic current as mentioned before, the contribution of the region S to the electric current can be also valuated.
The ballistic time is enough to accelerate a created carrier to $v_f$ effectively in $x$ direction.
Then, all what we need is the density of electric carriers $n(t_{bal})$, with which we can estimate the electric current as $j\sim en(t_{bal})v_f$ approximately\cite{gra_bal2, gra_bal3}.
$n(t_{bal})$ can be derived from the creation rate of the electric carriers $dn/dt$, which should correspond to the pair creation rate by the Schwinger mechanism, equation (\ref{eq:Sch_rate}).
As we mentioned before, in a state of an energy $E$ and a momentum $(p_x,p_y)$,
it takes the time $\Delta t$ for accomplishing the transition of the creation.
With the transition probability $T(p_x,p_y)$, the number of created particles per unit time is given by $T(p_x,p_y)/\Delta t$ for the state.
Consequently, the rate $dn/dt$ is given as follows,
\begin{eqnarray}
\frac{dn}{dt} &=& \frac{1}{(2\pi\hbar)^2}\int\int_{\rm S}dp_xdp_y\frac{T}{\Delta t}  \label{eq:int_S} \\
&=& \frac{e\varepsilon}{2(2\pi\hbar)^2}\int\int_{\rm S}dp_xdp_y\frac{p_x}{p_x^2+p_y^2} \nonumber \\
&\sim& \frac{e\varepsilon}{2(2\pi\hbar)^2}\int_0^{\frac{1}{2}\sqrt{\frac{e\varepsilon\hbar}{v_f}}}dp_x\int_{-\sqrt{\frac{e^2\varepsilon^2\hbar^2}{16v_f^2p_x^2}-p_x^2}}^{\sqrt{\frac{e^2\varepsilon^2\hbar^2}{16v_f^2p_x^2}-p_x^2}}dp_y\frac{p_x}{p_x^2+p_y^2}  \ \ \ (t_{bal}>>t_c)\nonumber \\
&=& \frac{e^{3/2}\varepsilon^{3/2}}{4\pi^2\hbar^{3/2}v_f^{1/2}} \int_0^1ds {\rm Arctan}\sqrt{\frac{1}{s^4}-1}
 = \frac{e^{3/2}\varepsilon^{3/2}}{4\pi^2\hbar^{3/2}v_f^{1/2}} \frac{B(\frac{1}{2}, \frac{3}{4})}{4},
\end{eqnarray}
where $B(m,n)$ represents the beta function
\footnote{
$B(m,n)=2\int_0^{\pi/2}({\rm sin}\theta)^{2m-1}({\rm cos}\theta)^{2n-1} d\theta$
}.
In the above approximation, the higher-order terms above $O(t_c/t_{bal})$ have been neglected because of $t_{bal}>>t_c$.
As $B(1/2,3/4)/4$ is about 0.6, this result roughly corresponds with the rate of the Schwinger mechanism, equation (\ref{eq:Sch_rate})\cite{Schwinger_add}. 
Note that, compared to this current by the Schwinger mechanism, the quasi-Ohmic current can be ignored in this case.

\section{Conclusion}
\label{sec:conclusion}
We have shown that a creation process of an electric carrier in graphene can be described by a weak-value formalism.
Although our weak-value model did not cover the entire physics, namely, the acceleration process, it was enough to show the feature of the electric currents for the cases of the quasi-Ohmic ($\propto \varepsilon$) and the Schwinger mechanism ($\propto \varepsilon^{3/2}$).
Our goal was not to make a strict estimation: not to settle in the value of the minimal conductivity.
However, our estimation of the currents approximately agrees with the earlier studies.
At the cost of a rigorous discussion, we have clarified the process to supply electric carriers in graphene related to virtual particles by the uncertainty relations.
In this sense, our approach is different from the earlier studies.
In the dynamical approach\cite{gra_bal2, gra_bal3}, the crossing time scale (\ref{eq:t_bal}) was derived, beyond which the perturbative treatment failed.
There, the Schwinger mechanism can be understood as a transmission picture with WKB approximation or be comprehensible in the context of the Landau-Zener transition\cite{Schwinger2, Schwinger3, Schwinger4}: 
considering the infinite past and future, the entire physics can be determined.
In contrast, our model describes the physics of the turning point of transition from a negative energy state to a positive one, i.e. the creation process itself, by which we have tried to valuate the entire current in graphene.
Note that the transition probability $T$ of equation (\ref{eq:trans_prob}) is irrelevant to the Landau-Zener transition probability.

A weak value was originally introduced by Aharonov, Albert and Vaidman as a result of weak measurements\cite{W1}.
According to a weak value, a pointer of a measurement apparatus is surely moved, although the weak value may take a strange value lying outside the eigenvalue spectrum.
While such physical effect to the pointer is one of the actual phenomena of a weak value, our model shows the case that a weak value emerges as an actual value of a physical quantity, following \cite{step}.
In addition, we have found the new insight in that a weak value of a group velocity makes a virtual particle in the uncertainty relations real.
Irrespective of a strange value, a weak value has often allowed us to treat a quantum particle as a classical one.
Interestingly, the figure depicted by (quantum) weak values, however, does not alway accord with the one of classical physics as shown in \cite{W_cl}.
A weak value seems to be simple, but not to be superficial.
Then, as we have seen, it needs to clarify how a weak value becomes effective in physics for understanding the meaning of the value.

\section*{Acknowledgements}
This work was supported by the Funding Program for World-Leading Innovative R \& D on Science and Technology (FIRST),
and JSPS Grant-in-Aid for Scientific Research(A) 25247068.

\end{document}